\documentclass[twocolumn]{aastex631}
\usepackage{amsmath}
\usepackage{xspace}

\usepackage{multirow}
\newcommand{\hst}{{\it HST}\xspace}
\newcommand{\swift}{{\it Swift}\xspace}
\newcommand{\cxo}{{\it Chandra}}
\newcommand{\xmm}{{\it XMM-Newton}\xspace}
\newcommand{\src}{R Aqr\xspace}

\newcommand{\revI}[1]{#1}
\newcommand{\revII}[1]{{#1}}

\begin{document}

\title{Front-row seat of the recent R Aqr periastron passage:\\
X-ray multi-epoch spectral and spatial analysis}

\correspondingauthor{Andrea Sacchi}
\author[0000-0002-7295-5661]{Andrea Sacchi}
\author[0000-0003-1769-9201]{Margarita Karovska}
\author[0000-0002-7868-1622]{John Raymond}
\author[0000-0002-3869-7996]{Vinay Kashyap}
\author[0000-0002-5115-1533]{Terrance J. Gaetz}
\affiliation{Center for Astrophysics $\vert$ Harvard \& Smithsonian, 60 Garden Street, Cambridge, MA 20138, USA}
\email{andrea.sacchi@cfa.harvard.edu}
\author{Warren Hack}
\affiliation{Space Telescope Science Institute, 3700 San Martin Drive,  Baltimore, MD 21218, USA}
\author[0000-0002-6745-4790]{Jamie Kennea}
\affiliation{Department of Astronomy and Astrophysics, The Pennsylvania State University, 525 Davey Lab, University Park, PA 16802, USA}
\author[0009-0006-5274-6439]{Nicholas Lee}
\affiliation{Center for Astrophysics $\vert$ Harvard \& Smithsonian, 60 Garden Street, Cambridge, MA 20138, USA}
\author[0000-0002-2564-3104]{Amy J Mioduszewski}
\author{Mark J Claussen}
\affiliation{National Radio Astronomy Observatory, P.O. Box O, 1003 Lopezville Road, Socorro, NM 87801, USA}







\begin{abstract}
We report on the X-ray spectral and spatial evolution of the Symbiotic star \src. Through a multi-epoch observational campaign performed with \cxo\ between 2017 and 2022, we study the X-ray emission of this binary system, composed of an evolved red giant star and a white dwarf (WD). This analysis is particularly timely as the WD approached the periastron in late 2018/early 2019, thus mass transfer, jet emission and outburst phenomena are to be expected.

Through detailed spectral analysis, we detect a significant rise in the soft X-ray ($0.5-2$ keV) emission of \src, likely linked to jet emission, followed by a decay towards the previous quiescent state. The hard X-ray emission ($5-8$ keV), is not immediately affected by the periastron passage; the hard component, after maintaining the same flux level between 2017 and 2021, rapidly decays after 2022. Possible explanations for this are a change in the reflection properties of the medium surrounding the binary, obscuration of the central region by material ejected during the periastron passage, or even the partial/complete destruction of the inner regions of the accretion disc surrounding the WD.

In addition to this activity in the central region, extended emission is also detected, likely linked to a hot spot in a pre-outburst-emitted jet, which can be observed moving away from the system's central region.
\end{abstract}



\section{Introduction} \label{sec:intro}

Symbiotic stars (SySts) are binary systems in which a compact object, usually a white dwarf (WD), accretes material from a red giant companion 
(see, e.g. \citealt{luna13} and references therein).
These systems are of crucial interest given that they are thought to be progenitors of asymmetric planetary nebulae (PN) and of at least a percentage of type Ia supernovae \revI{(see, e.g. \citealt{livio18} for a review)}.

X-ray observations are a powerful tool for studying SySts and in fact, the first classification scheme of SySts was proposed \citep{muerset97} based on {\em ROSAT} observations. The X-ray emission of SySts can be caused by different physical mechanisms, which often coexist in the same object: from nuclear burning reactions occurring on the compact object surface (producing extremely soft emission), to shocks caused by the interaction between the winds of the two companions, to hard X-ray emission produced by the hot accretion disc orbiting around the compact object.
The classification scheme relies on which of the spectral components of the SySt is dominant and it has been expanded beyond the {\em ROSAT} $0.5-2$ keV band, employing \swift\ data, by \citet{luna13}. SySts are dubbed $\alpha$-, $\beta$-, $\gamma$-, and $\delta$-type by increasing X-ray spectral hardness. Escaping this classification scheme, some peculiar SySt show X-ray spectra with coexisting soft and hard emissions, hence catalogued $\beta/\delta$-type.

\src\ is a SySt falling in this last category, composed of a WD and a Mira-type star with a pulsation period of 385 days. The two binary companions have a separation of about 10 AU and a suggested orbital period of about 42 years \citep{hollis97,gromadzki09}. Thanks to the system's closeness (at a distance of about 218~pc, it is \revI{one of} the closest known SySt \citealt{min14}) and spectacular multi-wavelength extended emission, characterized by a fascinating hourglass-shaped nebula with X-ray and radio bright jets and lobes, it has been the focus of dedicated observational campaigns (see e.g. \citealt{michalitsianos80,solf85,paresce94,kellogg01,kellogg07} and references therein). 

Recent multi-wavelength observations \citep{bujarrabal18,teyssier19,willson20}, strongly suggest that the \src\ WD companion has been approaching or is passing through the periastron since late 2018, several years earlier than predicted by the currently available orbit \citep{willson81}. The dimming of over 2 magnitudes of the optical light curve (V and B magnitudes, normally dominated by the Mira’s 7 mag amplitude variation), shows the signature of powerful recent mass ejection \revI{from the Mira}, resulting in a significant obscuration of the central region of the system, likely caused by a large dust cloud \citep{sankrit22} (see however \citealt{hinkle22}, where is suggested that the periastron passage might occur as late as 2024).

\begin{figure*}[ht!]
\centering
\includegraphics[width = \textwidth]{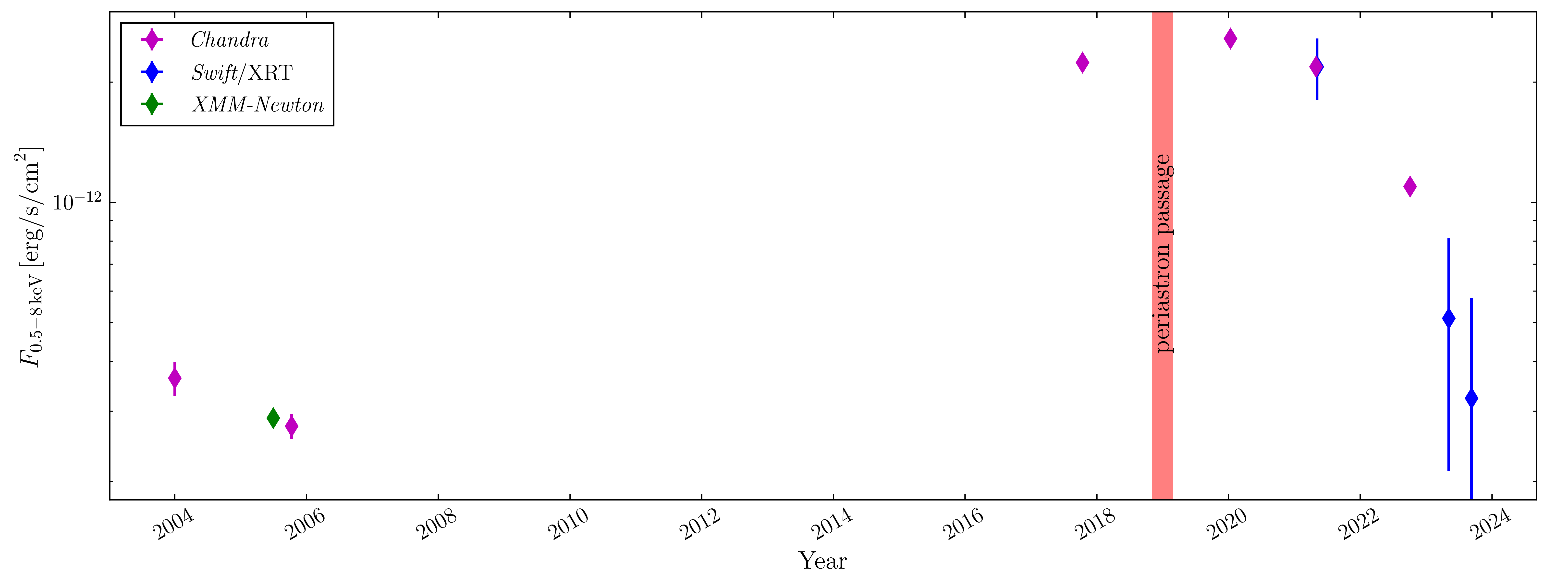}
\caption{X-ray long-term lightcurve of \src, on the $y$-axis the flux in the 
$0.5-8$ keV band and on the $x$-axis the epoch of the observation. The details of the observations used to build the plot are reported in Tab. \ref{tab:raqr_obs}. 
The red vertical bar highlights the beginning of the most recent periastron passage.
\label{fig:lc}}
\end{figure*}

This event represents a unique opportunity to study the system, as such events are linked to mass transfers, jet emission phenomena and dramatic changes in the observed properties of the system \citep{kafatos82}. This is also highlighted in Fig. \ref{fig:lc}, which shows the long-term X-ray lightcurve of \src: \cxo\ observations taken in 2017 before the WD periastron passage reveal the source being in a high state, with an X-ray luminosity more than an order of magnitude larger than that during 2005. Following the periastron passage, after a slight increase, the X-ray flux of \src\ decays rapidly.

In this paper, we focus on the results from the \cxo\ and \swift\ multi-epoch observations, analyzing in detail the innermost region of the source, where the activity linked to the periastron passage is concentrated. The data presented here have been obtained as part of an extended multiwavelength observational campaign (2018-ongoing), which includes space-based (\hst, \cxo, \swift) and ground-based (including VLA, and optical/IR photometric and spectroscopic) observations.

This paper is organized as follows: in Section \ref{sec:xray_data} we describe the X-ray dataset we employ; in Section \ref{sec:xspec} we present the X-ray spectral analysis of the central region; in Section \ref{sec:ext_emission} we report results of the analysis of the extended emission in the immediate proximity of the source (within 5"); in Section \ref{sec:disc} we discuss our results and, finally, in Section \ref{sec:conc}, we present our conclusions. \revI{All luminosities reported are computed assuming a distance of 218 pc \citep{min14}.}

\begin{table*}
\centering
\caption{Journal of the X-ray observations of R Aqr. Along with the instrument, \revI{observation id (ObsID)}, and epoch, we also report the exposure time and the flux of the source in the $0.5-8$ keV band. The observations presented in this paper are highlighted by a $\dag$ symbol.}
\label{tab:raqr_obs}
\begin{tabular}{ccccc}
\hline
Instrument & ObsID & epoch & $T_\textup{exp}$ (ks)& $\log F_{0.5-8\,{\rm keV}}$ \revI{(erg/s/cm$^2$)}\\
\hline
\cxo/ACIS & 4546 & 2003-12-31 & 36.5 & $-12.44\pm0.04$ \\
\xmm/pn & 04050101 & 2005-06-30 & 26.3 & $-12.54\pm0.02$ \\
\cxo/ACIS & 5438 & 2005-10-09 & 66.8 & $-12.56\pm0.03$ \\
\hline
\cxo/ACIS &  20809$^\dag$ & 2017-10-11 & 49.4 & \multirow{2}{*}{$-11.65\pm0.01$}\\
\cxo/ACIS & 19015$^\dag$ & 2017-10-13 & 75.6 & \\
\hline
\cxo/ACIS & 23108$^\dag$ & 2020-01-11 & 47.5 & $-11.59\pm0.01$ \\
\hline
\cxo/ACIS & 23325$^\dag$ & 2021-04-29 & 31.9 & \multirow{2}{*}{$-11.66\pm0.01$}\\
\cxo/ACIS & 24341$^\dag$ & 2021-04-30 & 35.6 & \\
\hline
\swift/XRT & 00014287001 & 2021-04-29 & 0.4 & \multirow{6}{*}{$-11.66\pm0.07$} \\
\swift/XRT & 00014287002 & 2021-04-29 & 0.6 & \\
\swift/XRT & 00014318003 & 2021-05-13 & 0.2 & \\
\swift/XRT & 00014318004 & 2021-05-13 & 1.0 & \\
\swift/XRT & 00014318005 & 2021-05-13 & 0.2 & \\
\swift/XRT & 00014318006 & 2021-05-13 & 1.0 & \\
\hline
\cxo/ACIS & 27333$^\dag$ & 2022-09-26 & 27.2 & \multirow{4}{*}{$-11.96\pm0.01$}\\
\cxo/ACIS & 27322$^\dag$ & 2022-10-03 & 18.8 & \\
\cxo/ACIS & 27467$^\dag$ & 2022-10-04 & 19.8 & \\
\cxo/ACIS & 27468$^\dag$ & 2022-10-09 & 14.9 & \\
\hline
\swift/XRT & 00011564042 & 2023-05-06 & 0.1 & \multirow{4}{*}{$-12.30\pm0.20$} \\
\swift/XRT & 00011564043 & 2023-05-06 & 1.5 & \\
\swift/XRT & 00011564046 & 2023-05-07 & 0.2 & \\
\swift/XRT & 00011564047 & 2023-05-07 & 1.5 & \\
\hline
\swift/XRT & 00011564050 & 2023-09-08 & 3.0 & \multirow{2}{*}{$-12.50\pm0.25$}\\
\swift/XRT & 00011564051 & 2023-09-10 & 2.8 & \\
\hline
\end{tabular}
\end{table*}

\section{X-ray data set} \label{sec:xray_data}
Table \ref{tab:raqr_obs} summarizes the details of all the \cxo, \xmm, and the most recent \swift\ observations of \src. For each observation the instrument, ObsID, exposure time and flux are reported. 

All \cxo\ data were reprocessed and reduced with the Chandra Interactive Analysis of Observations software package (\texttt{CIAO}, v.4.12; \citealt{fruscione06}) and the \texttt{CALDB} 4.9.0 release of the calibration files.

Spectra were extracted from the individual observations and then merged using the \texttt{CIAO} tool \texttt{combine\_spectra} when taken close to one other, namely, we merged the two 2017, the two 2021 and the four 2022 spectra.

Similarly, the images for each epoch were obtained by merging the single observations with the \texttt{dmmerge} tool, after correcting for the small shift between individual observations, identified by tracing the brightest pixels, with the \texttt{wcs\_update} and \texttt{reproject\_events} tools.

Spectra of the \cxo\ observations taken in 2003 and 2005 were extracted from a $1"$-radius circular region centred on the source's position, determined in every observation using the \texttt{ds9} tool \texttt{centroid}. Fluxes have been derived with a model similar to the ones described in Section \ref{sec:xspec}, which details the spectral analysis performed on the observation involved in the periastron passage (marked
in Tab. \ref{tab:raqr_obs}).

\xmm\ data were processed with the Science Analysis Software (\texttt{SAS}, v.20.0; \citealt{gabriel04}) and calibration files obtained on 20 February 2023, following standard SAS routines: periods of high background were eliminated by inspecting the lightcurve of the full observation in the $0.2-12$ keV energy band. The EPIC/pn spectrum was extracted from a circular region of 15" radius centred on the source position and the background from a 40" circular, free-of-sources region, on the same detector chip. The flux was obtained with a model similar to the one described in Section \ref{sec:xspec}.

\swift/XRT data were reprocessed using the dedicated tool \texttt{xrtpipeline} (v.0.13.7) and merged using the \texttt{extractor} routine. Count rates were obtained with the \texttt{ximage} tool \texttt{sosta} and converted into fluxes adopting the best fitting model of the closest available \cxo\ observations.

\section{Central region spectral analysis} \label{sec:xspec}

As all the activity linked to the most recent periastron passage occurs in the immediate vicinity of the central source, X-ray spectra of this region were extracted from a $1"$-radius circular region centred on the source's position (cf. Fig. \ref{fig:multi_ep}), which encloses about 95\% of the point spread function (PSF). 

Background spectra were extracted from a $15"$-radius source-free, circular region, placed at about $35"$ distance, southeast with respect to the central region.

Spectra were extracted with the \texttt{specextract} tool, rebinned to have at least one count per bin and fed into the spectral fitting package \texttt{XSPEC} \citep{arnaud96} version 12.12.1, where the Cash statistic \citep{cash79} implementation in \texttt{XSPEC} (cstat) was employed. Any further rebinning, adopted to create the figures presented, is purely graphical.

We fit the spectra extracted from the central region, considering counts in the $0.5-8$ keV band, with a multi-component model following the approach adopted for other SySts (e.g., CH\,Cyg, \citealt{mukai07,karovska10}). We employed two thermal plasma components (\texttt{vapec} in \texttt{XSPEC}) to describe the soft emission and a heavily absorbed component to account for the hard emission, plus a Gaussian emission line, with energy fixed at 6.4\,keV, to reproduce the observed neutral (or near-neutral) Fe K$\alpha$. The heavily absorbed thermal component was able to account for the rest of the Fe-K complex, the 6.7\,keV line (He-like ``triplet'') and the 6.97\,keV line (Ly$\alpha$-like line).

The mid-band emission ($2-5$ keV), has been addressed, in literature, with several physically-motivated models, including, but not limited to, thermal plasmas, reflection by a neutral medium, and reflection by an ionized medium \citep{mukai07,luna19,toala23}. We tested all these models, but, while any model could reproduce the data, given the ``narrowness" of the band of interest and the quality of our data, none of the chosen model parameters could be constrained, hence we decided to adopt a simple power law model.

Finally, all the absorbers adopted were modelled with the \texttt{TBabs} absorption model \citep{wilms00}.

\begin{figure*}[ht!]
\centering
\includegraphics[width = \textwidth]{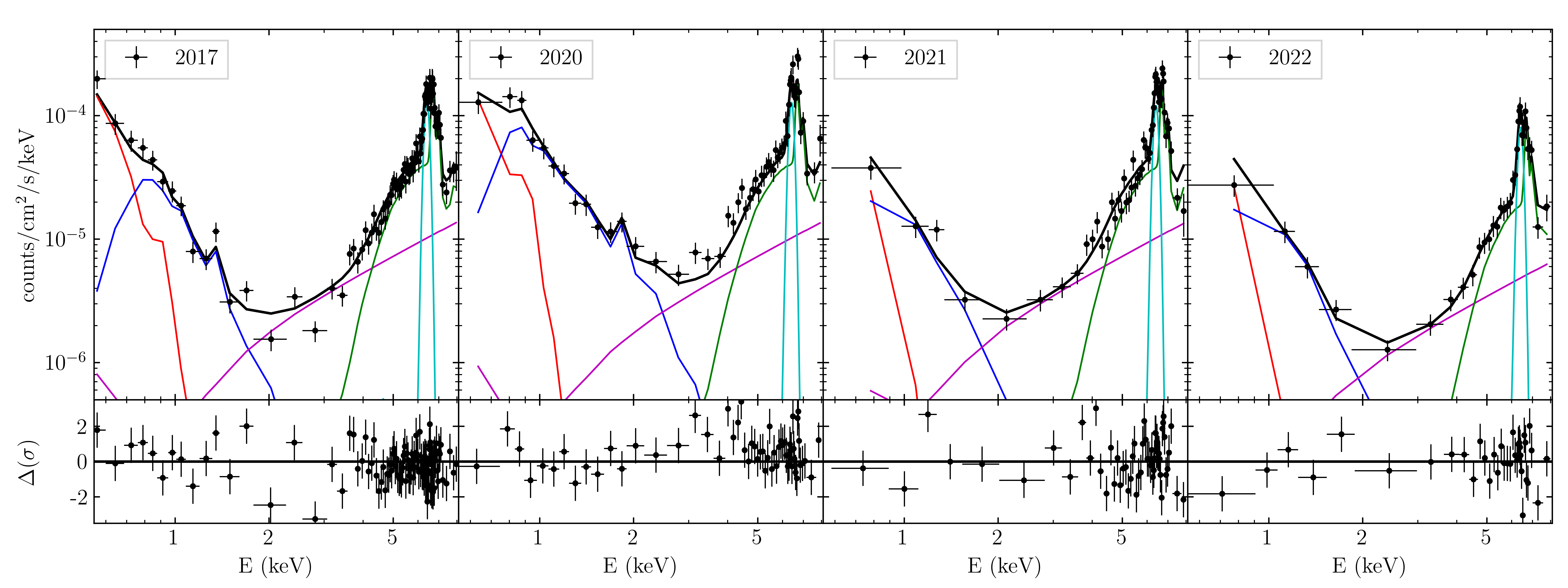}
\caption{
Observed X-ray spectra and best-fit models for each epoch. 
From left to right, 2017, 2020, 2021, and 2022 X-ray spectra (upper panels) and residuals (lower panels) of the central region. Data and the full model are in black, the single component in coloured solid lines: red, blue and green for the thermal plasmas, the power law is in magenta and the Gaussian line is in cyan. 
\label{fig:spec_mosaic}}
\end{figure*}

\begin{figure}[ht!]
\centering
\includegraphics[width = 0.5\textwidth]{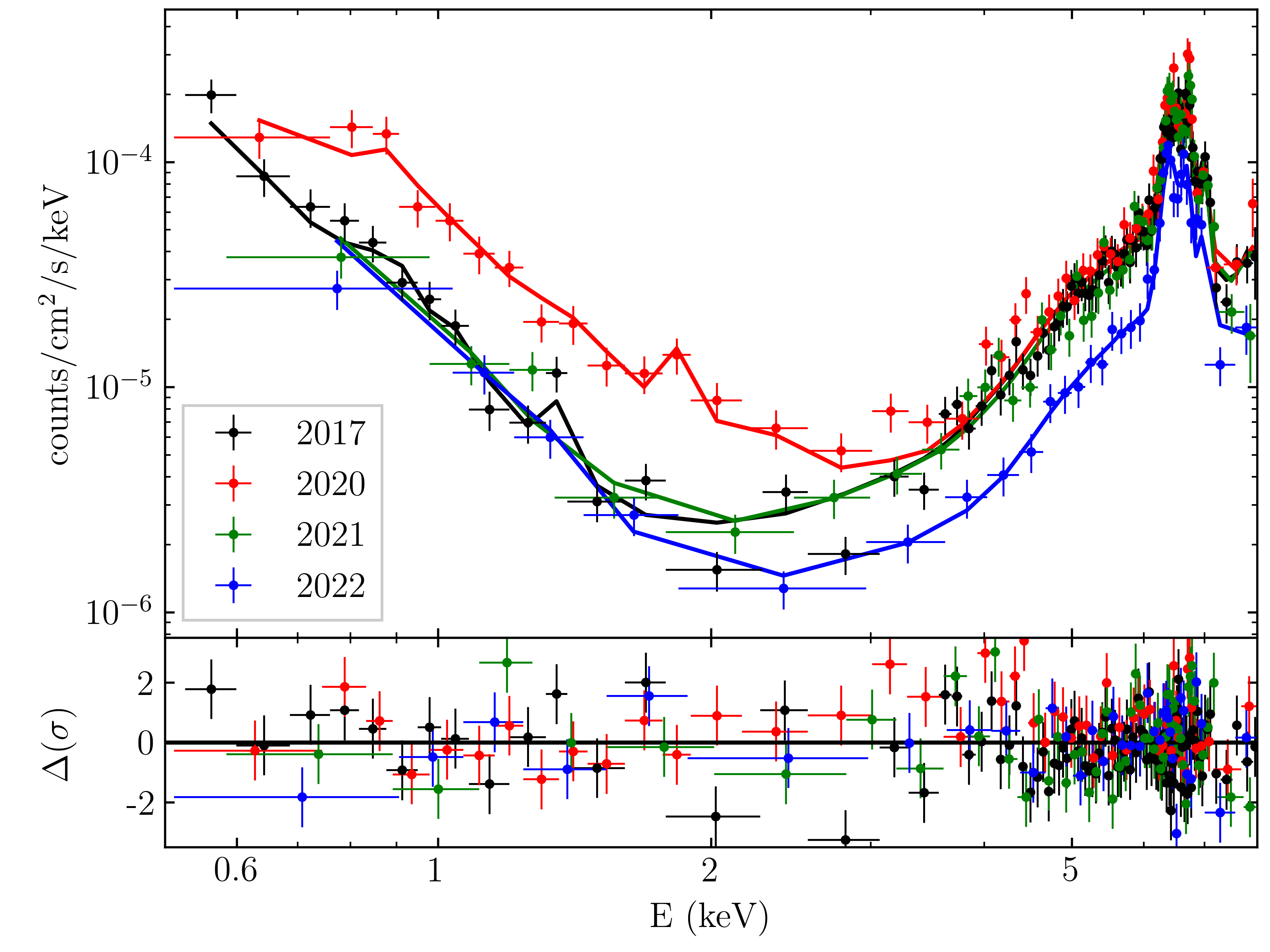}
\caption{X-ray spectra and residuals of the central region.
\label{fig:spec_soft}}
\end{figure}

The final model we adopted is hence:

\begin{equation*}
\begin{split}
    &\texttt{TBabs}_1\times(\texttt{vapec}_1+\texttt{vapec}_2+\texttt{powerlaw})\\
    +&\texttt{TBabs}_2\times(\texttt{vapec}_3+\texttt{Gauss}).
\end{split}
\end{equation*}
The spectra of the four epochs were fitted simultaneously in \texttt{Xspec}, \revI{minimizing the C-statistic differences between the model convolved with the \cxo\ response matrix and the observed count rate spectra.} We obtained an acceptable statistic ($C/\nu=2179.80/1587\approx1.37$, where $C$ is the value of statistic and $\nu$ the number of degrees of freedom) for the parameters reported in Tab. \ref{tab:c_spec}.  The multi-epoch spectra, along with the different components of the model, are shown in Fig. \ref{fig:spec_mosaic}, while Fig. \ref{fig:spec_soft} shows the comparison of the spectra over the different epochs.

\begin{table*}
\centering
\caption{Spectral parameters of the best-fitting model of the central region. Values are reported with $1\sigma$ uncertainties. Parameters indicated by (f) and (t) are, respectively, fixed and tied together during the fitting procedure.}
\label{tab:c_spec}
\begin{tabular}{ll|cccc}
\hline
 & & \multicolumn{4}{c}{epoch}\\
component & parameter & 2017 & 2020 & 2021 & 2022\\
\hline
\texttt{TBabs}$_1$ &$N_\textup{H}\,(10^{22}\,{\rm atoms/cm}^2)$ & \multicolumn{4}{c}{$0.95_{-0.12}^{+0.14}$ (t)} \\
\hline
\multirow{2}{*}{\texttt{vapec}$_1$} & $kT\,({\rm keV})$ & $0.057\pm0.004$ (t) & $0.062\pm0.004$ & $0.057\pm0.004$ (t) & $0.057\pm0.004$ (t)\\
& Norm & \multicolumn{4}{c}{$5.4_{-3.0}^{+10.7}$}\\
\hline
\multirow{2}{*}{\texttt{vapec}$_2$} & $kT\,({\rm keV})$ & $0.30\pm0.01$ (t) & $0.65\pm0.06$ & $0.30\pm0.01$ (t) & $0.30\pm0.01$ (t)\\
& Norm & \multicolumn{4}{c}{$(2.7\pm0.5)\times10^{-4}$ (t)} \\
\hline
& $F_\textup{0.5-2 keV}$ (erg/cm$^2$/s) & $(6.6\pm0.3)\times10^{-14}$ (t) & $(1.67\pm0.04)\times10^{-13}$ & $(6.6\pm0.3)\times10^{-14}$ (t) & $(6.5\pm0.4)\times10^{-14}$\\ 
\hline
\multirow{2}{*}{\texttt{powerlaw}} & $\Gamma$ & \multicolumn{4}{c}{$-1.3\pm0.2$ (t)} \\
& Norm & \multicolumn{3}{c}{$(8.6\pm0.2)\times10^{-7}$ (t)} & $(4.1\pm0.1)\times10^{-7}$\\
\hline
& $F_\textup{2-5 keV}$ (erg/cm$^2$/s) & $(1.55\pm0.03)\times10^{-13}$ & $(1.67\pm0.03)\times10^{-13}$ & $(1.43\pm0.04)\times10^{-13}$ & $(6.1\pm0.3)\times10^{-14}$\\ 
\hline
\texttt{TBabs}$_2$ & $N_\textup{H}\,(10^{22}\,{\rm atoms/cm}^2)$ & \multicolumn{2}{c}{$110\pm5$ (t)} & $116\pm5$ & $134\pm12$\\
\hline
\multirow{2}{*}{\texttt{vapec}$_3$} & $kT\,({\rm keV})$ & \multicolumn{4}{c}{$6.2\pm0.3$ (t)}\\
& Norm & \multicolumn{3}{c}{$(1.5\pm0.1)\times10^{-2}$ (t)} & $(0.91\pm0.01)\times10^{-2}$ \\
\hline
& $F_\textup{5-8 keV}$ (erg/cm$^2$/s) & $(2.05\pm0.02)\times10^{-12}$ & $(2.05\pm0.02)\times10^{-12}$ & $(1.94\pm0.02)\times10^{-12}$ & $(1.01\pm0.03)\times10^{-12}$\\ 
\hline
\multirow{3}{*}{\texttt{Gauss}} & $E\,({\rm keV})$ & \multicolumn{4}{c}{6.4 (f)}\\
& $\sigma\,({\rm keV})$ & \multicolumn{4}{c}{$0.10\pm0.01$ (t)} \\
& Norm & \multicolumn{3}{c}{$(1.4\pm0.1)\times10^{-4}$ (t)} & $(1.1\pm0.1)\times10^{-4}$ \\
\hline
& $F_\textup{Gauss}$ (erg/cm$^2$/s) & $(3.7\pm0.2)\times10^{-13}$ & $(3.7\pm0.2)\times10^{-13}$ & $(3.5\pm0.2)\times10^{-12}$ & $(2.2\pm0.2)\times10^{-13}$ \\ 
\hline
\end{tabular}
\end{table*}

\begin{figure}[ht!]
\centering
\includegraphics[width =0.5 \textwidth]{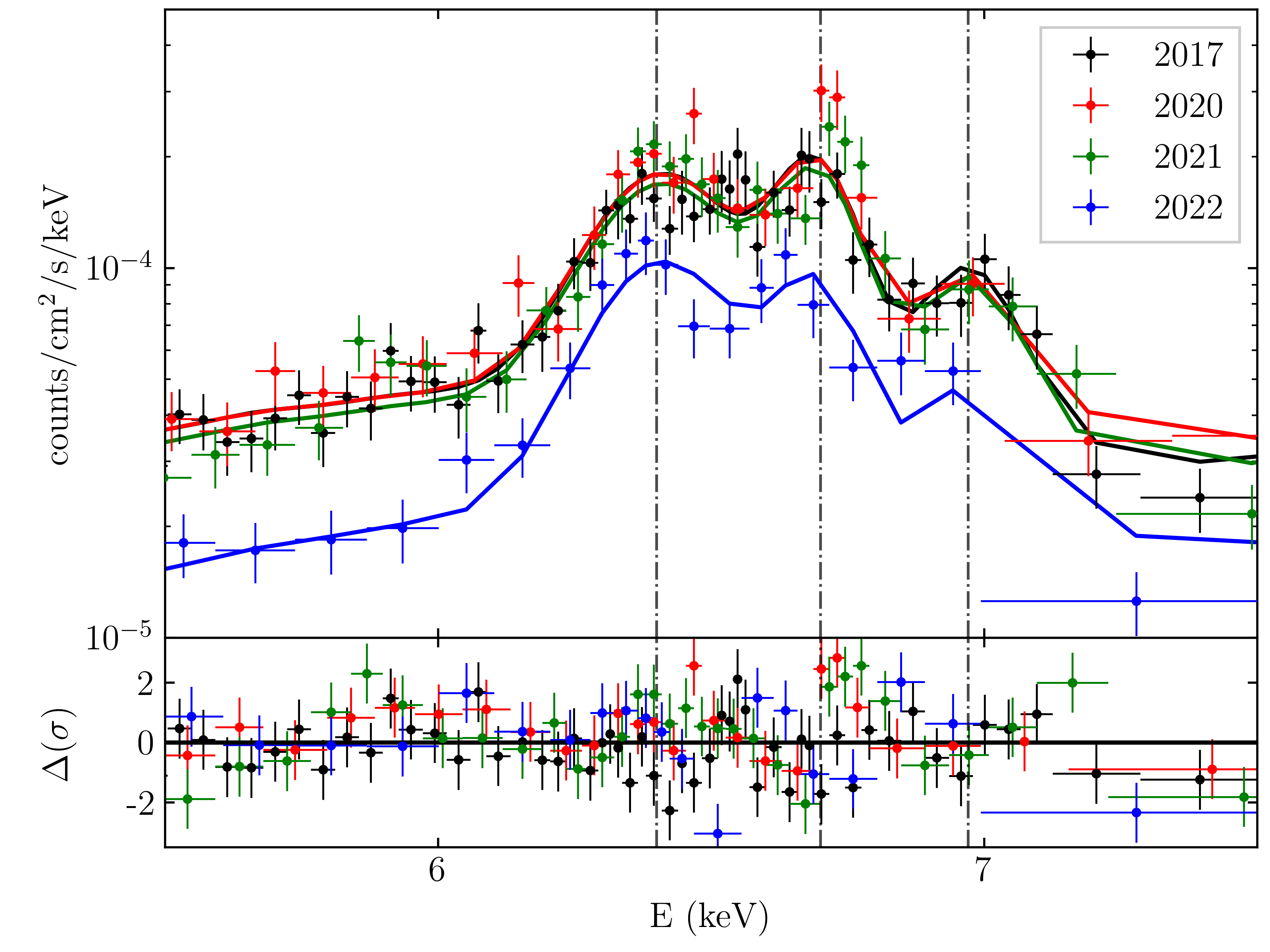}
\caption{X-ray spectra and residuals of the central region, zoom on the Iron complex. Vertical dash-dotted lines highlight the positions of the three Iron lines.
\label{fig:spec_hard}}
\end{figure}

It is evident from the fits that the soft- and hard-energy portions of the \src\ spectrum evolve in very different ways. 

In the soft-energy band, compared to the 2017 observation, \src\ shows an increase in flux, of about a factor of 2.5 in 2020. It then drops down again and does not change significantly in 2021 and 2022. We can attribute the increase in flux to variations in the temperatures of the two soft thermal components: the softest component (\texttt{vapec$_1$}) increases marginally by $\approx$10\% in 2020 while the hotter component increases by $\gtrsim{2}\times$ from 0.3~keV in 2017 to 0.65~keV in 2020 before returning to 2017 levels subsequently. 

In the hard-energy band, \src\ shows little to no evolution until 2022, when the column density of the absorber increases by $\gtrsim$15\%, and the normalizations of both the high-temperature thermal plasma component \texttt{vapec$_2$}) and the Gaussian line (\texttt{Gauss}) decrease by $\approx$20-40\%.

We employed the \texttt{vapec} model in order to allow the elemental abundances to deviate from Solar. Unfortunately, the quality of our spectra does not allow us to perform such discrimination: no statistically significant variation from solar abundances is detectable.

Furthermore, the thermal-plasma component used to model the hard emission can also account for two of the three observed emission lines in the Iron complex: the Fe\textsc{xxv} at $6.7$ keV and the Fe\textsc{xxvi} at $6.97$ keV. However, to investigate the evolution of these two emission lines with time, we substituted the standard thermal plasma model with a no-line thermal plasma one (\texttt{nlapec} in \texttt{Xspec}) and modelled the emission lines with two additional Gaussians with fixed energies. This allows us to independently investigate the evolution of the continuum emission and one of the lines. In this case, too, we are not able to detect any statistically significant evolution in the line emission, except a drop of about a factor of 1.5 in the 2022 observation, consistent with the one observed for the continuum. A zoom-in on the Iron complex of the X-ray spectra of \src\ is shown in Fig. \ref{fig:spec_hard}.\\

\section{Extended emission analysis}\label{sec:ext_emission}

Here we present a detailed analysis of the extended emission of \src.  Although not directly related to the most recent periastron passage, this analysis is prompted by the jet emission observed in UV, optical, and radio wavelengths as well as in previous X-ray observations \citep[see, e.g.,][]{kellogg07,toala22}.

\begin{figure*}[ht!]
\centering
\includegraphics[width = \textwidth]{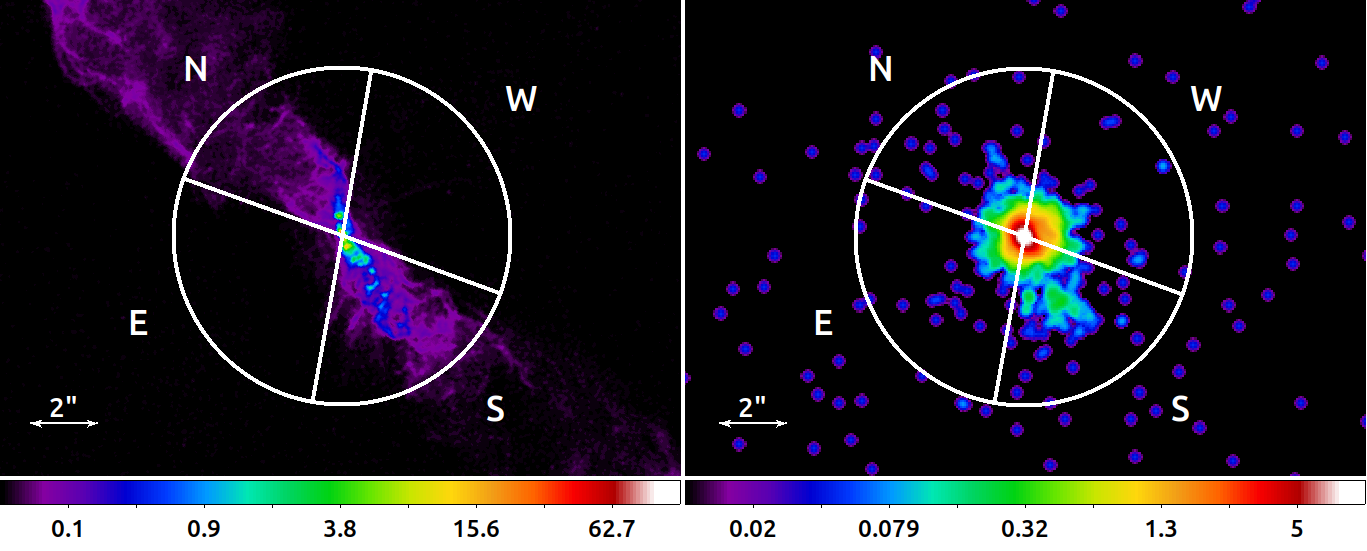}
\caption{The quadrants we use to analyze the extended emission in \src, superposed on the [O\textsc{ii}] \revI{\hst\ image taken in 2021 (on the left) and X-ray image taken in 2020 (\revII{on the right})}. The X-ray image has been created at a binning of $1/8$ of native pixel size and is then smoothed with a Gaussian of $\sigma=4$~binned image pixels. \revII{The \hst\ image is in units of electrons/s, the \cxo\ image is in units of counts.} \label{fig:secs}}
\end{figure*}

Based on a comparison with the \hst\ [O\textsc{ii}] image, we draw four quarter circles (see Fig. \ref{fig:secs}) centred on \src's central source position and with a 5" radius. Sectors (N)orth and (S)outh have been chosen to include the jet locations, while sectors (E)ast and (W)est cover the jet-free regions.

To investigate the presence of extended emission, we then simulate the \revI{PSF} for each epoch, and in each sector compare the radial profile of the observation and the simulated PSF.

We follow the approach described by \citet{falcao23} and exploited the \cxo\ science threads\footnote{\url{https://cxc.cfa.harvard.edu/ciao/threads/}}. We first employ \texttt{ChaRT}\footnote{\url{https://cxc.harvard.edu/ciao/PSFs/chart2/}}, the \cxo\ PSF ray tracer simulator, to take into account the central region spectrum (described in Section \ref{sec:xspec}) the off-axis angle and the aspect solution of each observation. Then we project the generated rays on the detector with \texttt{MARX}\footnote{\url{https://cxc.cfa.harvard.edu/ciao/threads/marx_sim/}}. 

\begin{figure*}[ht!]
\centering
\includegraphics[width = \textwidth]{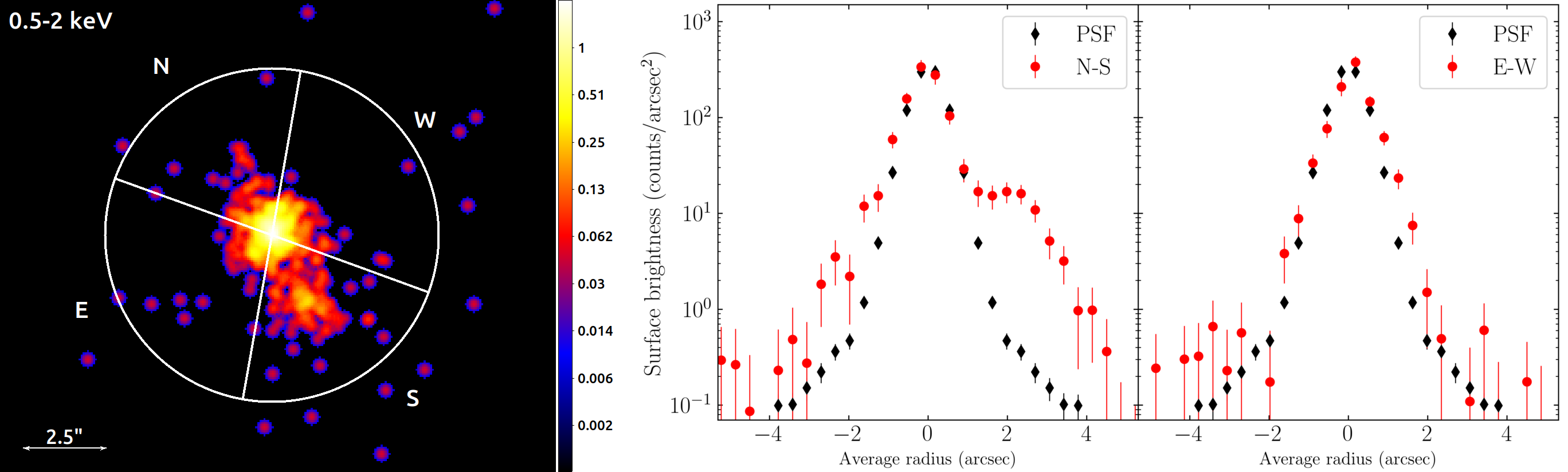}
\includegraphics[width = \textwidth]{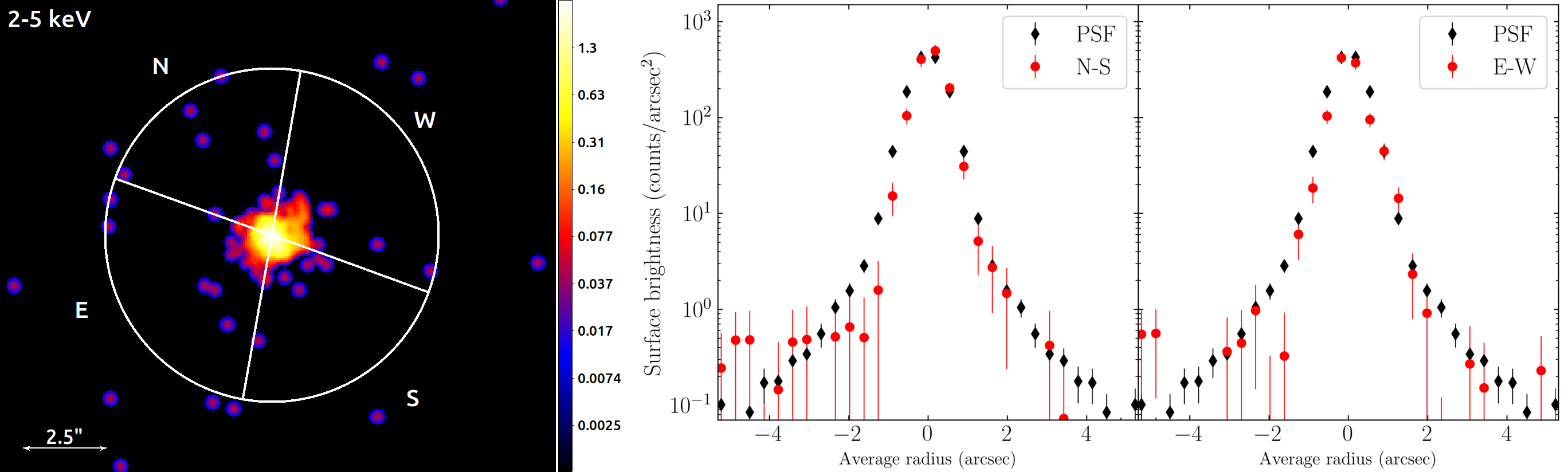}
\includegraphics[width = \textwidth]{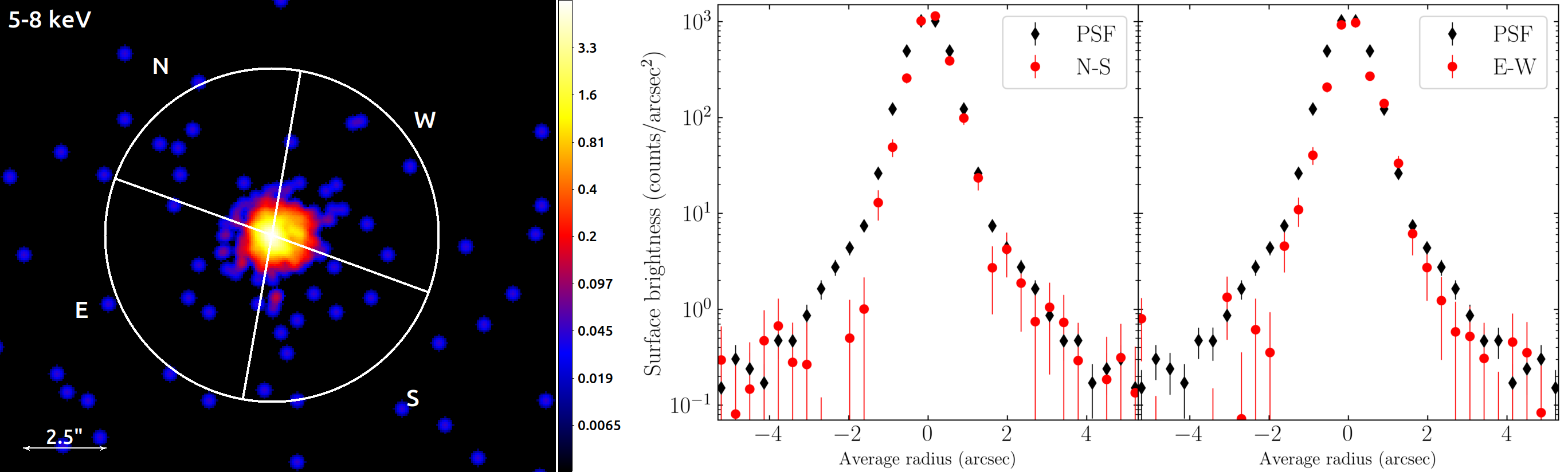}
\caption{For the 2020 observation, from top to bottom, soft ($0.5-2$ keV), mid ($2-5$ keV), and hard ($5-8$ keV) band. On the left, is the X-ray image in each band (smoothed with a Gaussian kernel with a 4-pixel radius). In the middle the comparison between the radial profiles extracted from the North (negative radius values) and South (positive radius values) sectors, within red dots the original data background is subtracted and in black diamonds the PSF. On the right the same plots but for the East and West sectors. \revII{All images are in units of counts.}
\label{fig:psf}}
\end{figure*}

For each epoch, we then construct the radial profile in the sectors described above, combining the E-W and N-S quadrants for both the observation and the simulated PSF, in three energy bands: $0.5-2$ keV, $2-5$ keV, and $5-8$ keV. Figure \ref{fig:psf} illustrates the comparison between the radial profiles of the observed (red points) and simulated PSF (black points) for the 2020 observation, of the three energy bands, going along both the jet direction (N-S) and the cross-jet direction (E-W).

We do not detect differences, within the S/N limits, between the observed and simulated PSF along the cross-jet direction (East and West sectors) in any epoch. The small deviations that appear to exist in the tails are consistent with Poisson fluctuations. In contrast, we find faint soft emission in the North sector, in 2017 and 2020, getting fainter in 2021 and 2022. Unfortunately, there are insufficient excess counts here to carry out spectral fits. We find significant extended emission in the soft energy band in the South quadrant. This excess evolves over time, and we describe the details and implications of this excess in Sec. \ref{sec:blob_dyn} below.

\begin{figure*}[ht!]
\centering
\includegraphics[width = 0.75\textwidth]{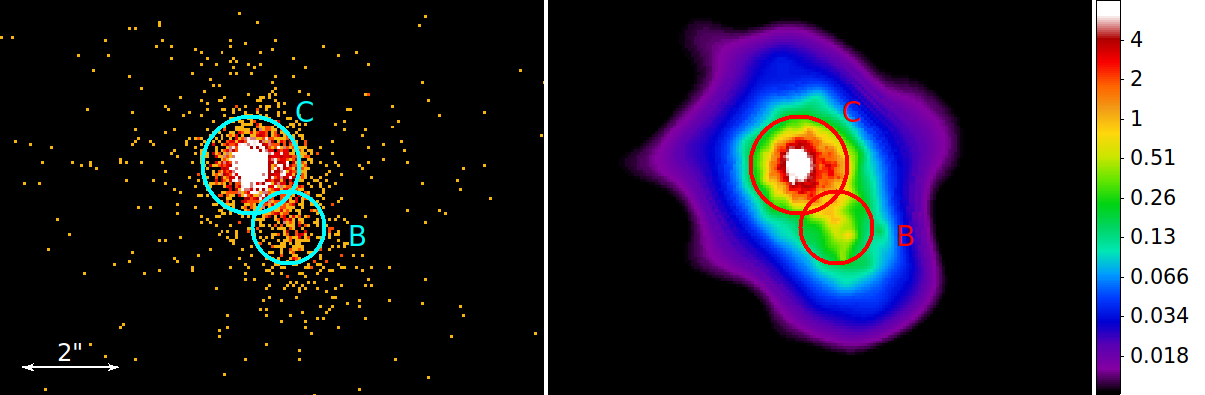}
\includegraphics[width = 0.75\textwidth]{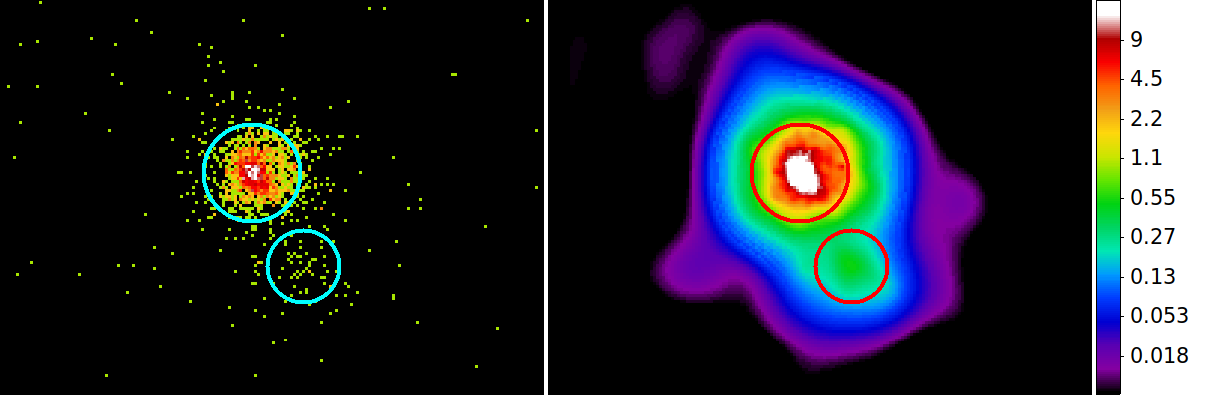}
\includegraphics[width = 0.75\textwidth]{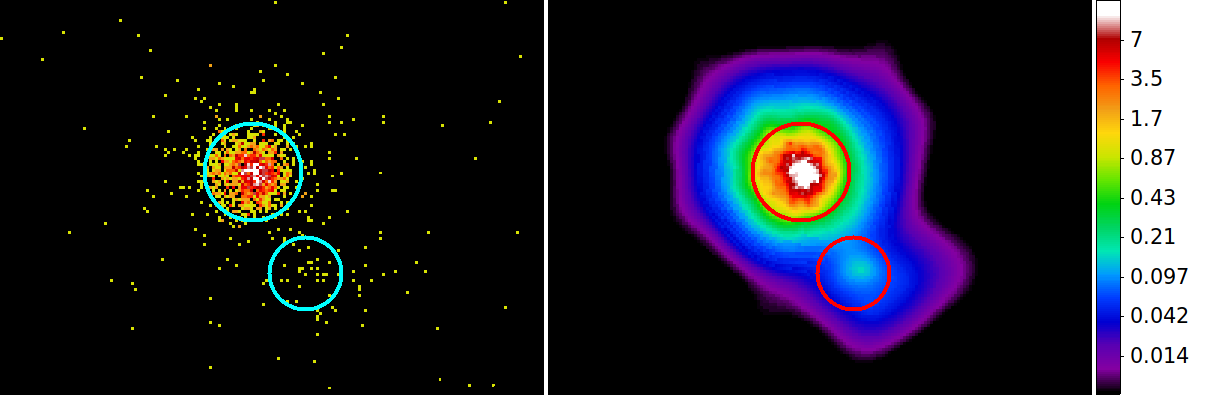}
\includegraphics[width = 0.75\textwidth]{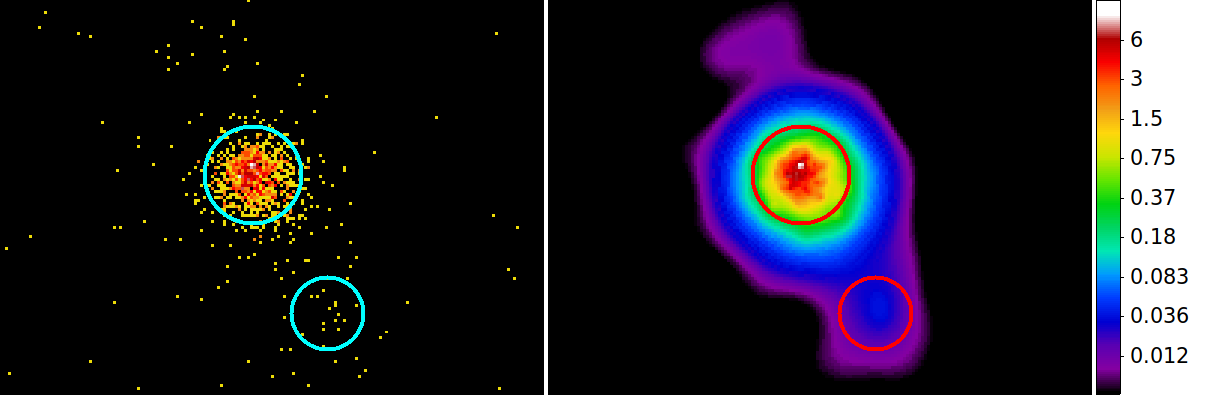}
\includegraphics[width = 0.75\textwidth]{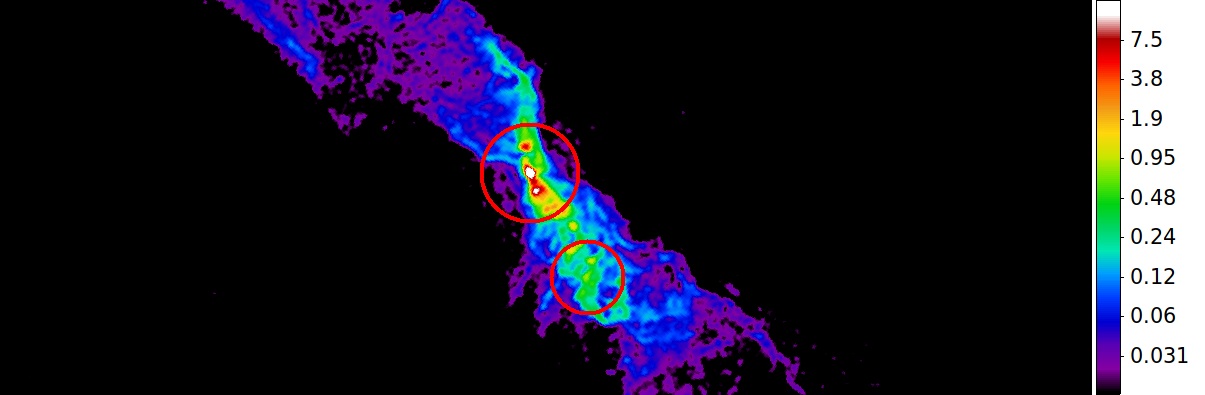}
\caption{From top to bottom, 2017, 2020, 2021, and 2022 X-ray images in the soft band ($0.5-2$ keV). On the left is the ``raw'' image, and on the right are the adaptively smoothed images, with 10 counts under the kernel, $0\farcs5$ minimum radius and 8'' maximum radius. The circles show the central region(C) and blob (B) regions. The last row shows for comparison the \hst\ image obtained in 2021 with the F373N, $\lambda\,3730$ \AA, [O\textsc{ii}] line filter. \revII{All \cxo\ images are in units of counts. The \hst\ image is in units of electrons/s.}
\label{fig:multi_ep}}
\end{figure*}

Upon visual inspection, the extended emission appears to be located in a roughly circular region, hereafter dubbed ``blob". The blob is superposed to the jet location and, across the different epochs, it appears to be moving away from the central region. This is clearly shown by Fig. \ref{fig:multi_ep}, which presents the X-ray images of \src\ over the different epochs, with highlighted the central region (labelled C) and the blob (labelled B).

The signal-to-noise ratio in the blob is high enough that spectral analysis can be performed (see Sec. \ref{sec:blob} below). 

\subsection{Blob spectral analysis}\label{sec:blob}

X-ray spectra of the blob were extracted from a $0\farcs75$-radius circular region, centred on the brightest pixel of the blob itself. We judge that this choice represents the best compromise between avoiding contamination from the central region and maximizing the signal-to-noise ratio for the blob.

Given that extended emission has only been detected in the soft band, only photons between $0.5$ and 2 keV were considered.

\begin{figure}[ht!]
\centering
\includegraphics[width =0.5 \textwidth]{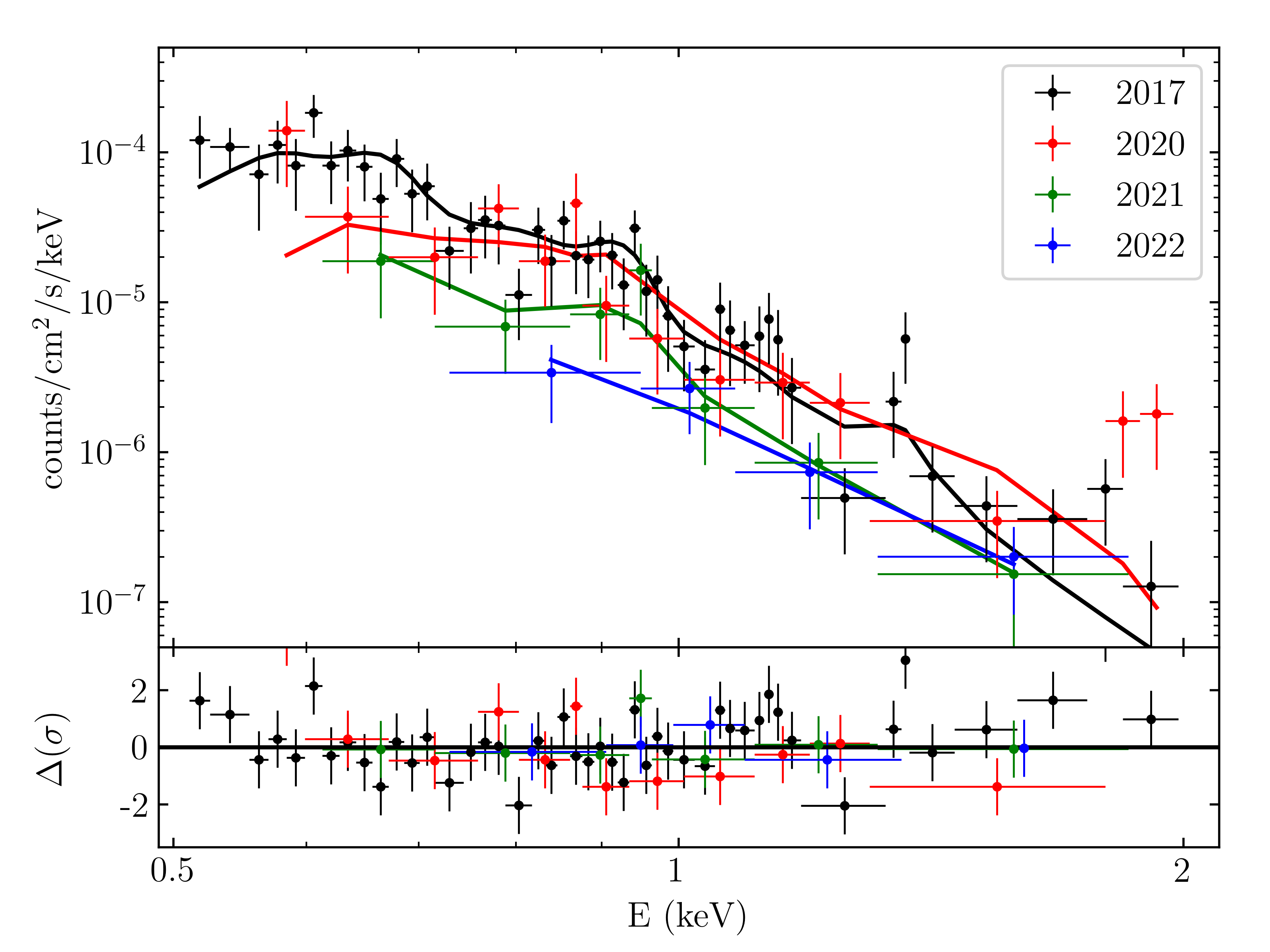}
\caption{X-ray spectra and residuals of the blob.
\label{fig:blob_spec}}
\end{figure}

The blob spectrum in different epochs was modelled with just one absorbed thermal plasma model. The best-fit parameters and flux in the $0.5-2$ keV band are reported in Tab. \ref{tab:blob_spec} and the spectra in the four epochs are shown in Fig. \ref{fig:blob_spec}. For the 2017 spectrum, contamination from the PSF of the central source cannot be excluded, but, given the distance of the blob from the central region, it does not contribute to more than 8\%.

\begin{table}
\centering
\caption{Best-fit spectral parameters of the blob. Values are reported with $1\sigma$ uncertainties. \label{tab:blob_spec}}\begin{tabular}{cccc}
\hline
\multirow{2}{*}{epoch} & $N_\textup{H}$ & $kT$ & $F_{0.5-2\,{\rm keV}}$\\
 & ($10^{22}\,{\rm atoms/cm}^2$) & (keV) & ($10^{-14}\,{\rm erg/s/cm}^2$)\\
\hline
2017 & $0.28\pm0.07$ & $0.18\pm0.01$ & $3.9\pm0.1$\\
2020 & $0.55\pm0.12$ & $0.20\pm0.02$ & $1.8\pm0.2$\\
2021 & $0.44\pm0.16$ & $0.16\pm0.02$ & $1.4\pm0.3$\\
2022 & $0.88\pm0.32$ & $0.17\pm0.02$ & $0.3\pm0.2$\\
\end{tabular}
\end{table}

Neither the temperature of the thermal plasma nor the column density of the absorber shows appreciable trends across the different epochs; the former being around 200 eV and the latter varying between 3 and $9\times10^{21}$ atoms/cm$^{2}$ but with large error bars. The only appreciable systematic effect is the decrease in flux, which, from the first to the last observation, drops by an order of magnitude over five years.

\section{Discussion} \label{sec:disc}

\src\ is a highly variable symbiotic star.  It is of considerable interest now because of the recent periastron passage of the WD companion which occurred in late 2018/early 2019.  This event has been covered in detail with an extensive multiwavelength observational effort, with observations spanning from late 2017 to late 2022.  In this work, we focus on the X-ray spectral and spatial emission of \src, based on nine \cxo-ACIS observations.\\

\subsection{Blob dynamic and evolution}\label{sec:blob_dyn}
The most notable feature revealed by the X-ray spatial analysis is a hot spot in the south/southwest direction, spatially superposed on the southern jet. This hot spot {\sl blob} has a roughly circular shape in the non-deconvolved image and can be seen moving away from the central region while fading, as shown in Fig. \ref{fig:multi_ep}. This, coupled with its characteristic X-ray spectral shape, best-modelled by a thermal plasma with a temperature $kT\lesssim200$ eV, makes it compatible with being the result of shocks and collisions heating up material along the jet, in a similar fashion to what has been observed in other SySts \citep{karovska10}.

\begin{figure}[ht!]
\centering
\includegraphics[width =0.5 \textwidth]{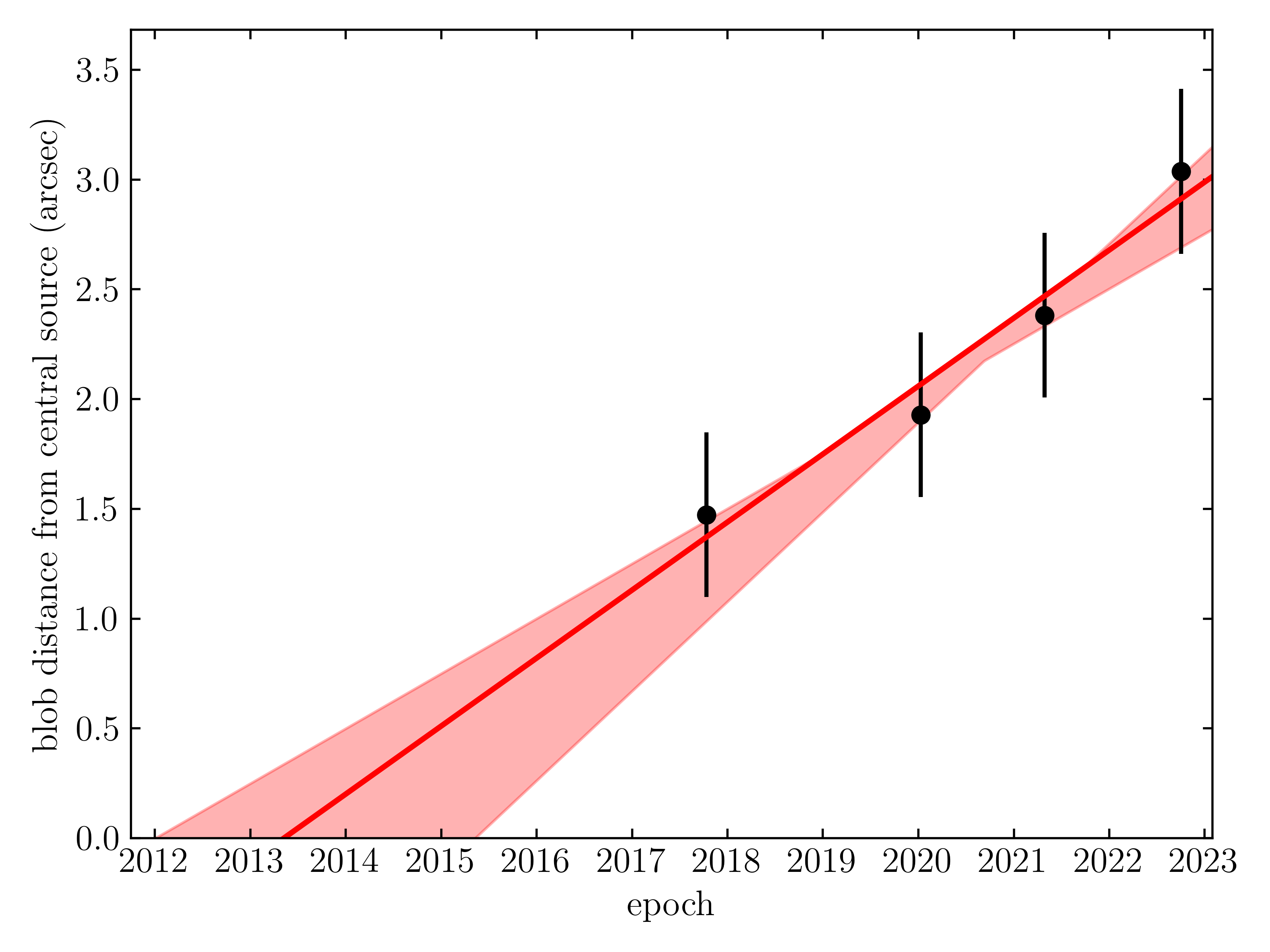}
\caption{Projected distance between the blob and the central region in the four epochs. Shaded in red, the region comprised between the line extrapolated using only the first and last three epochs. The solid red line shows the extrapolation using the four epochs.  
\label{fig:blob_pos}}
\end{figure}

The blob's projected velocity is $\approx250$ km/s if computed using the position inferred from the three oldest epochs, and $\approx400$ km/s if using the three most recent ones (blob positions in the different epochs are shown in Fig. \ref{fig:blob_pos}). These velocities are considerably larger (by a factor of 3-5) than the ones inferred from the motion of the knots observed in the jet at optical wavelengths \citep{melnikov18,huang23}, however both these features exhibit acceleration with increasing distance from the central region. This suggests that, rather than the actual motion of the jet material, we are observing the propagation of a shockwave along the jet: a shock speed of $\sim$400 km s$^{-1}$ is consistent with the observed temperatures near 0.2 keV. Extrapolating the blob position backwards places its emergence between 2012 and 2015\revI{, as shown in Fig. \ref{fig:blob_pos}}.\\

\subsection{\src\ in the context of accreting WD}

In the usual scenario for cataclysmic variables, accretion at low rates in quiescent dwarf novae produces hot, thermal emission at a temperature of $5-10$ keV from a vertically thick boundary layer where the accretion disk contacts the WD and dissipates the energy of Keplerian rotation.  At higher accretion rates, above roughly $10^{16} ~\rm g~s^{-1}$, in dwarf novae in outburst and in nova-like variables, the higher densities allow the boundary layer to cool and collapse into a thin disk that radiates the accretion energy as a quasi-blackbody with a temperature of $0.02-0.05$ keV. The blackbody component dominates the luminosity, but some high-temperature emission is produced as well, perhaps from the low-density material at the surface of the accretion disk \citep{patterson85a, patterson85b}.  At high accretion rates, the disk or boundary layer also produces a strong, weakly collimated wind at speeds around 3000 $\rm km~s^{-1}$ with a kinetic energy of order 10\% of the total luminosity \citep{drew87, mauche87, long02}.  There is evidence of a jet in the dwarf nova SS Cyg when it makes a transition between states, but jets in CVs seem to be rare \citep{coppejans20}.  Jets in SySts seem to be more common \citep{brocksopp04}, either because interaction with the red giant wind makes them more visible, or because jets are created when the red giant wind collimates the wind of a WD accreting at a high rate. Within this picture, one would expect a large increase in accretion rate as the WD approaches periastron, perhaps leading to a change from a low accretion to a high accretion state.

We fit the X-ray spectrum of \src\ with 5 components;  a) a thermal emission component with $kT\sim$0.06 keV and luminosity $\sim 4\times 10^{35}\,\rm erg~s^{-1}$\revI{, \texttt{vapec}$_1$ in our model and in Tab. \ref{tab:c_spec}}, b) a thermal emission component with $kT=0.3-0.6$ keV and luminosity $\sim 4 \times 10^{32}\,\rm erg~s^{-1}$, \revI{\texttt{vapec}$_2$}, c) a thermal emission component with $kT\sim6$ keV and luminosity $\sim 4 \times 10^{32}\,\rm erg~s^{-1}$, \revI{\texttt{vapec}$_3$}, d) a Fe K$\alpha$ emission line with luminosity $\sim 2 \times 10^{30}\,\rm erg~s^{-1}$, \revI{\texttt{Gauss}}, and e) a power law component with $\Gamma \sim -1.3$, \revI{\texttt{powerlaw} in our model}.  We discuss them individually.


a) The lowest temperature thermal component, which we have fit with a thermal spectrum, could also be fit with a blackbody of a similar temperature.  In either case, the luminosity estimate of $\sim 4\times 10^{35} \rm ~erg~s^{-1}$ is somewhat uncertain because of the significant absorption, but the nominal value would require an accretion rate of about $2 \times 10^{18} \rm ~g~s^{-1}$ on a 1 solar mass WD. This is in the range of CVs in outburst and \revI{the estimated luminosity is an order of magnitude larger than the luminosity of the $\sim$ 40,000 K component that \citet{burgarella92} invoked to explain the optical emission from the jet of R Aqr (although their model was meant for the low accretion state). One would expect that all emission components from the boundary layer should be subject to the same absorbing column density, while the fitted $N_\textup{H}$ values for the soft component and the 6 keV component are $0.95 \times 10^{22}$ and $120 \times 10^{22} \rm ~cm^{-2}$, respectively. This discrepancy could be reconciled if interpreted as the outcome of obscuration by the Mira's wind. Under reasonable assumptions about the Mira's mass loss, the binary separation, and the ionizing photon luminosity, the region dominated by the Mira's neutral wind has a conical shape \citep{seaquist84,nussbaumer87}. In this geometry, different components can be seen through different portions of the wind, resulting in different amounts of absorption. This might also suggest that the soft component arises from shocks in the red giant winds rather than from the boundary layer, and hence is seen through a lower-density column of absorbing material.}

b) The $0.3-0.6$ keV component must be thermal emission since the Mg\textsc{xi} line near 1350 eV is apparent in some of the spectra. That is most plausibly identified with shocks in the red giant wind driven by the jet or wind from the WD. The luminosity is only 0.1\% the luminosity of the cooler component.  That would suggest $\sim$ 1\% the power of the wind is dissipated in slower shocks of around 600 $\rm km~s^{-1}$\footnote{\revI{The temperature behind a shock is about $0.012\,V_{100}^2$ keV, where $V_{100}$ is the shock speed in units of 100 km/s \citep{cox72}.}} driven into the denser gas of the red giant wind. The emission from these shocks would not be isothermal but would span a range of temperatures.

c) The 6 keV emission must indeed be thermal emission because it shows Fe\textsc{xxv}/Fe\textsc{xxvi} emission lines. However, in the context of accreting WDs, it is not expected to be isothermal. \citet{perna03} analyzed the HETG spectrum of the quiescent dwarf nova WX Hyi and, while the ratio of the Fe\textsc{xxv} line to Fe\textsc{xxvi} suggested a temperature around 6 keV, an isothermal model failed to match most of the other lines, apparently because a range of temperatures was present.  Not only did an isothermal model provide a poor match, but other models that had been proposed, including a hot boundary layer \citep{narayan93}, a coronal siphon \citep{meyer94}, a hot corona \citep{mahasena99} and a hot settling flow \citep{medvedev02} all failed in one way or another, indicating that important physics is missing from even these more sophisticated models. Only in some cases, a cooling flow model can match accreting WD spectra \citep{mukai03, luna07}. Nevertheless, accretion onto the WD is the most plausible source of the 6 keV component and an accretion rate of order $10^{15} \rm ~g~s^{-1}$ is required. However, the details of the spectral shape are not understood, so the value of N$_H$ derived from the fits should be treated with caution.

d) The 6.4\,keV Fe~K$\alpha$ line could be produced by fluorescence of cool gas illuminated by the 6 keV component. It could arise from the surface of the WD, the accretion disk, or the red giant surface or wind. While it is most commonly observed in magnetic CVs \citep{mukai15}, it is also seen in nonmagnetic ones \citep{rana06}.  Since its luminosity is less than 1\% that of the 6 keV component and the red giant wind provides a large solid angle to absorb X-rays, its presence is not surprising.

e) The fit includes a power law component primarily to fill in the spectral range between the soft and hard components ($2-4$ keV). Power law X-ray emission is rare in accreting WDs, and given that both hard and soft components are multithermal plasmas approximated as isothermal ones, it is likely that the emission at intermediate energies is actually additional thermal emission. An alternative explanation for the emission in this energy band involves the reflection of hard X-rays by the material surrounding the system. This model has been recently employed to reproduce the X-ray spectrum of CH Cyg, another SySts \citep{toala23}.

\subsection{Effects of the recent periastron passage}

The effects of the most recent WD periastron passage are to be looked for in the innermost vicinities of the central regions, roughly within 1", which is the region where our spectral analysis is focused. 

Right before the periastron passage, in late 2017, \src\ is in a high-flux state, with a typical $\beta/\delta$-type SySt spectrum, showing both soft and hard X-ray emission, discussed above. This might be already related to the incoming periastron passage: the WD would start attracting more material as the binary separation decreases, as also suggested by hydrodynamical simulations \citep{deValBorro17}.

Following the periastron passage, in 2020, the soft flux of \src\ increases by a factor of $\approx2.5$. We modelled this with a higher-temperature plasma \revI{for the component described above in point (b)}, with respect to the 2017, pre-periastron passage observation. We interpret the soft-emission of \src\ and its flux increase in 2020 as a result of jet launching activities. This is based on three considerations: a) this portion of the X-ray spectrum is usually linked to jet interaction with the surrounding medium, b) the blob, likewise linked to the jet propagation, shares a very similar spectrum, and c) during the periastron passage \revI{enhanced accretion state}, some sort of new jet emission is expected.

The increase in the soft-band flux extends to the $2-4$ keV band and, as highlighted above, we lack a satisfactory explanation for this component. To exclude, however, the possibility that this flux increase is linked to pile-up effects, given the significant rise observed in the softer emission, we performed a round of simulation with the \cxo\ simulator \texttt{MARX} \citep{davis12}, which confirmed that pile-up effects are not responsible for this mid-band emission, prompting for further theoretical and modelling effort in order to explain its nature. 

During this period the hard-band emission is unaffected, the flux remaining stable within the uncertainties.

After the 2020 increase in flux, in 2021 soft- and mid-band emission returns to pre-periastron levels. In addition, a small but statistically significant increase in the column density of the hard-band absorber is seen, suggesting that the periastron passage also starts to affect the hard band in this epoch. 

In the 2022 September/October observations, while the soft band is not affected anymore and maintains, within uncertainties, the level of the previous observation, the hard-band emission shows a decrease in the flux of a factor of $\approx1.5$. This is caused by an increase in the column density of the absorber and a drop in the normalization of both the thermal plasma and the Gaussian line at $6.4$ keV.

After 2022, no other high-quality X-ray spectrum of \src\ is available. However, short \swift/XRT observations, carried out in May and September 2023, allow us to assess the flux level of the source: as clearly shown in Fig. \ref{fig:lc} the flux of \src\ is steadily decreasing, and it will likely reach the 2005 level soon.

Different mechanisms might account for the spectral changes observed in the hard X-ray band. One possible scenario invokes obscuration: the hard X-ray emission, coming from the innermost region of the accretion disc surrounding the WD would be blocked by the material ejected. Supporting this scenario is the change observed in the column density of the absorber of the hard component. In this scenario, however, it is difficult to understand the delay between the response of the soft- and hard-band emission. A possible way around this would be invoking obscuration by dust. Mira-type stars are known to emit large amounts of dust \citep{khouri18}, and ALMA observation detected abundant dust in the circumbinary region of \src, including the orbital plane \citep{ramstedt18,sankrit22}. The delayed response in the hard X-ray band could then be understood in terms of the time needed for the dust to condense in the orbital plane after the powerful outburst occurred during the periastron passage. 

Another viable explanation involves reflection. \citet{toala23} demonstrated that a reflection model can explain the emission of $\beta/\delta$-type SySts in the mid-X-ray band ($2-4$ keV) as well as account for a significant fraction of the hard-X-ray emission, including the fluorescent line at 6.4 keV. A change in the physical properties and/or geometry of the reflection medium could easily explain the drop in flux in the hard band observed after 2021.

Finally, in a more drastic scenario, the drop in hard-X-ray flux could be caused by the partial (or even complete) destruction of the innermost portion of the accretion disc surrounding the WD. This could explain the observed and still persisting decay in flux, and it would represent a catastrophic failure of the self-regulating mechanism between mass-accretion and outflows which is thought to be governing accretion discs in SySts \citep{lucy20}.

\section{Conclusions} \label{sec:conc}
In this paper, we present the X-ray spectral and spatial analysis of the innermost region of \src, a Symbiotic Star composed of a Mira-type star and a WD, which recently underwent its periastron passage. The periastron passage, for this kind of system, is a particularly interesting and dramatic event due to the increase in mass transfer and close interaction between the winds of the two companions.

Owing to 9 \cxo\ observations, and several \swift\ visits, we monitored \src\ from late 2017 to mid-2023 in order to optimally cover the periastron passage, which started likely in late 2018/early 2019.

Following the periastron passage we detected an increase in the soft X-ray emission of the source, which we tentatively interpret as due to the formation of a new jet and consequent shocks between the material ejected and the material surrounding the system.

The hard-X-ray emission instead shows a delayed response: it does not show any significant variation up to 2021, and then drops steadily up to the most recent \swift\ observations, carried out in May and September 2023. We propose several scenarios to explain this delayed drop in the flux above 4 keV, including obscuration by dusty material, a change in the physical properties and/or geometry of a reflection medium surrounding the system, or a complete destruction of the innermost portion of the accretion disc.

Further removed from the central region and not linked with the recent periastron passage, we also find extended soft emission, from a roughly circular region in the deconvolved images, spatially coincident with the jet location. This hot spot gets fainter and moves away from the central region with time. We interpret this emission as a result of shock within the jet, or between the jet and the surrounding material.   

The results from the X-ray spectroscopy presented in this paper will be combined with the results from the multiwavelength observational campaign since 2017 in an upcoming dedicated publication.

\begin{acknowledgments}
We thank the anonymous referee for their useful comments. We are grateful to AAVSO and its dedicated observers, and to the ARAS group (Francois Teyssier) for the observing campaigns of \src\ since 2017, obtaining timely photometry and spectra, which were critical for planning the \hst\ and \cxo\ observations. We are also grateful to the \swift\ ToO program which allowed timely observations coordinated with the \cxo\ and \hst\ observations. AS thanks Caroline Huang, Rodolfo Montez Jr. and Jesús Toalá for insightful discussions.
AS acknowledges support from NASA \cxo\ Award Number DD9-20111X. MK acknowledges support provided via the NASA \cxo\ grants GO7-18021X, DD9-20111X, GO1-22027X, and GO3-24014X and NASA/ESA Hubble Space Telescope (which is operated by the Association of Universities for Research in Astronomy, Inc..) grants HST-GO-16055.001-A, HST-GO-16312.001-A, and GO3-24014X. VK, TJG, and NL were supported by the NASA Contract NAS8-03060 to the Chandra X-ray Center.
\revI{This paper employs a list of \cxo\ datasets, obtained by the {\em Chandra X-ray Observatory}, contained in~\dataset[DOI: 10.25574/cdc.180]{https://doi.org/10.25574/cdc.180}.}
\end{acknowledgments}

%







\bibliography{biblio}{}
\bibliographystyle{aasjournal}

\end{document}